\documentstyle[12pt]{article}
\def\thebibliography#1{\leftline{\it References}\list
  {[\arabic{enumi}]}{\settowidth\labelwidth{[#1]}\leftmargin\labelwidth
    \advance\leftmargin\labelsep
    \usecounter{enumi}}
    \def\newblock{\hskip .11em plus .33em minus .07em}
    \sloppy\clubpenalty4000\widowpenalty4000}

\begin{document}

\rightline{SU--4240--588}
\rightline{UNITU--THEP--23/1994}
\rightline{October 1994}
\rightline{hep-ph/9410320}
\vskip2cm
\centerline{\Large\bf Particle Conjugation and the
$1/N_C$ Corrections to $g_A$}
\vskip2cm
\centerline{J.\ Schechter$^{a}$ and H.\ Weigel$^{a,b}$}
\vskip1cm
\centerline{$^a$Department of Physics, Syracuse University}
\centerline{Syracuse, NY 13244--1130, USA}
\vskip0.5cm
\centerline{$^b$Institute for Theoretical Physics,
T\"ubingen University}
\centerline{D-72076 T\"ubingen, Germany}
\vskip 3cm
\baselineskip=20 true pt
\centerline{\bf ABSTRACT}
\vskip .25cm

We impose the requirement that the isovector axial vector current
for the soliton sector of the chiral quark model transforms correctly
under particle conjugation. This forces us to choose an otherwise
arbitrary ordering of collective space operators in such a way that the
next--to--leading $1/N_C$ correction to $g_A$ vanishes.

\vfill\eject

\normalsize
\baselineskip=22 true pt
Recently, the very interesting observation has been made
\cite{wa93}--\cite{wa94} that the too small value of the neutron beta
decay constant $g_A$ predicted in many chiral soliton models might be
dramatically improved by including subleading $1/N_C$ corrections.
These calculations, however, require one to make a certain choice about
the ordering of two collective space operators which appear in the
expression for the current. An ambiguity arises when one replaces
classical objects by quantum operators while doing the collective
quantization. In the present note we show that particle conjugation
invariance can provide useful operator ordering information. It implies
that the next--to--leading $1/N_C$ correction to $g_A$ should vanish in
the context of collectively quantizing the static soliton configuration.
We will use the chiral quark model \cite{kr84} but the calculation is
essentially identical to those in other models of this type
\cite{wa93}--\cite{wa94}. More detailed aspects as well as a similar
study for the isovector part of the magnetic moment of the nucleon will
be presented elsewhere \cite{sch94}.

In chiral models it is useful to employ the
non--linear parametrization $U={\rm exp}\left(i\mbox{\boldmath $\tau$}
\cdot\mbox{\boldmath $\pi$}/f_\pi\right)$ of the pion field,
$\mbox{\boldmath $\pi$}$. The static soliton configuration is described
in terms of the hedgehog {\it ansatz}
\begin{eqnarray}
U_0(\mbox{\boldmath $r$})={\rm exp}\Big[i
\mbox{\boldmath $\tau$}\cdot
{\hat{\mbox{\boldmath $r$}}}F(r)\Big],
\label{hedge}
\end{eqnarray}
where the radial function $F(r)$ is obtained by minimizing the static
energy functional while maintaining definite boundary conditions. The
projection of the soliton onto physical states is accomplished by
introducing \cite{ad83} time dependent collective coordinates
$A(t)\in SU(2)$ parametrizing the isospin (and/or spin) orientation
of the hedgehog
\begin{eqnarray}
U(\mbox{\boldmath $r$},t)=
A(t)U_0(\mbox{\boldmath $r$})A^{\dag}(t).
\label{rothedge}
\end{eqnarray}
Due to the hedgehog structure of $U_0$ a rotation in space is
equivalent to a {\it right} transformation of $A(t)$, {\it i.e.}
the spin operator represents the {\it right} generator in the
collective space
\begin{eqnarray}
[\mbox{\boldmath $J$},A]=A\frac{\mbox{\boldmath $\tau$}}{2}.
\label{comrules}
\end{eqnarray}
The angular velocity $\mbox{\boldmath $\Omega$}$, defined by
\begin{eqnarray}
\frac{i}{2}
\mbox{\boldmath $\tau$}\cdot\mbox{\boldmath $\Omega$}=
A^{\dag}\frac{\partial}{\partial t}A,
\end{eqnarray}
measures the {\it right} transformation at the classical level. When
quantizing $A(t)$ canonically, the spin is identified as the momentum
conjugate to $\mbox{\boldmath $\Omega$}$
\begin{eqnarray}
\mbox{\boldmath $J$}=\frac{\partial L}
{\partial \mbox{\boldmath $\Omega$}}=
\alpha^2\mbox{\boldmath $\Omega$},
\label{defspin}
\end{eqnarray}
where $L=L(A,\mbox{\boldmath $\Omega$})$ denotes the collective
Lagrangian. The moment of inertia, $\alpha^2$, is a
functional of the chiral angle, $F(r)$ and is of order $N_C$.

In the chiral quark model \cite{kr84} one considers a Dirac spinor field
in the background field $U$
\begin{eqnarray}
{\cal L}_{\rm q}= {\overline \Psi}
\left(i{\partial \hskip -0.5em /}
-mU^{\gamma_5}\right)\Psi,
\label{lq}
\end{eqnarray}
where $m$ parametrizes the quark--pion coupling. The contribution
of the quarks to the static energy functional is given by
\begin{eqnarray}
E_{\rm q}^{\rm cl}=N_C{\rm sgn}(B)\epsilon_{\rm val},
\label{eqcl}
\end{eqnarray}
where $\epsilon_{\rm val}$ denotes the lowest (in magnitude) eigenvalue
of the Dirac Hamiltonian
\begin{eqnarray}
h(F)={\mbox {\boldmath $\alpha$}} \cdot {\mbox{\boldmath $p$}} +
\beta m \Big[{\rm cos}F(r) + i\gamma_5{\mbox{\boldmath $\tau$}}
\cdot{\hat{\mbox{\boldmath $r$}}}\ {\rm sin}F(r)\Big].
\label{hstat}
\end{eqnarray}
In (\ref{eqcl}) the sign of the baryon number carried by the soliton is
included as an additional factor to accommodate the hole interpretation
of the Dirac theory. Next the collective coordinates are taken to be
time dependent. Transforming to the flavor rotating frame
$\Psi\rightarrow A\Psi$ adds the Coriolis term
$\mbox{\boldmath $\tau$}\cdot\mbox{\boldmath $\Omega$}/2$ as a
perturbation to (\ref{hstat}). Then $\epsilon_{\rm val}$ is changed
at second order yielding the quark part of the collective Lagrangian
\begin{eqnarray}
L_{\rm q}=-E_{\rm q}^{\rm cl}
+\frac{1}{2}{\rm sgn}(B)\alpha^2_{\rm q}
{\mbox{\boldmath $\Omega$}}^2\ , \qquad
\alpha^2_{\rm q}=\frac{N_C}{2} \sum_{\mu\ne{\rm val}}
\frac{\left|\langle\mu|\tau_3|{\rm val}\rangle\right|^2}
{\epsilon_{\rm val}-\epsilon_\mu},
\label{alquark}
\end{eqnarray}
while the wave--function acquires a so--called cranking
correction to become \cite{in54}
\begin{eqnarray}
\Psi_{\rm crank}=A(t)\left\{\Psi_{\rm val}
+\frac{1}{2}\sum_{\mu\ne{\rm val}}\Psi_\mu
\frac{\langle\mu|\mbox{\boldmath $\tau$}\cdot
\mbox{\boldmath $\Omega$}|{\rm val}\rangle}
{\epsilon_{\rm val}-\epsilon_\mu}\right\}.
\label{delval}
\end{eqnarray}

Now, the contribution of the {\it quarks} to the isovector
axial vector current $A_i^a$ is
obtained by substituting (\ref{delval}) into
${\rm sgn}(B)N_C\bar{\Psi}\gamma_i\gamma_5(\tau^a/2)\Psi$. The factor
${\rm sgn}(B)$ has to be carried along because this current is
a Noether current of the functional (\ref{alquark}).
Furthermore $\mbox{\boldmath $\Omega$}$ is replaced by
$\mbox{\boldmath $J$}/\alpha^2$ according to (\ref{defspin}) yielding
the quark part of  the axial current operator in the space of the
collective coordinates
\begin{eqnarray}
A_i^{({\rm q})a} &=& {\rm sgn}(B)N_C D_{ab}
\Psi^{\dag}_{\rm val}\sigma_i\frac{\tau^b}{2}\Psi_{\rm val}
\label{delcur} \\
&&\hspace{1cm}
+{\rm sgn}(B)\frac{N_C}{2\alpha^2}
\left[(1-\xi)J_jD_{ab}+\xi  D_{ab}J_j\right]
\sum_{\mu\ne{\rm val}}
\frac{\langle{\rm val}|\tau_j|\mu\rangle}
{\epsilon_{\rm val}-\epsilon_\mu}
\Psi^{\dag}_\mu\sigma_i\frac{\tau^b}{2}\Psi_{\rm val}.
\nonumber \\
&&\hspace{1cm}
+{\rm sgn}(B)\frac{N_C}{2\alpha^2}
\left[(1-\eta)D_{ab}J_j+\eta J_jD_{ab}\right]
\sum_{\mu\ne{\rm val}}
\frac{\langle\mu|\tau_j|{\rm val}\rangle}
{\epsilon_{\rm val}-\epsilon_\mu}
\Psi^{\dag}_{\rm val}\sigma_i\frac{\tau^b}{2}\Psi_\mu.
\nonumber
\end{eqnarray}
Here $\xi$ and $\eta$ denote {\it a priori} undetermined parameters
labeling possible orderings between
$D_{ab}=\frac{1}{2}\ {\rm tr}\left(\tau_aA\tau_bA^{\dag}\right)$
and $\mbox{\boldmath $J$}$. Their presence reflects the ambiguity
occurring when going from the classical to the quantum description.
As $\alpha^2$ is of the order $N_C$ the ambiguous terms are subleading.

We note that the eigenvalues $\epsilon_\mu$ are degenerate with respect
to the grand spin ({\it i.e.} total spin plus isospin) projection,
$M_\mu$. The associated sum introduces a kind of reduced matrix element
$T_\mu$
\begin{eqnarray}
\sum_{M_\mu}\langle{\rm val}|\sigma_3\tau_b|\mu,M_\mu\rangle
\langle\mu,M_\mu|\tau_j|{\rm val}\rangle&=&
iT_\mu\epsilon_{3bj}.
\label{sproj}
\end{eqnarray}
An explicit calculation shows that $T_\mu$ is real \cite{sch94}. The
current (\ref{delcur}) then yields the quark contribution to
$g_A$
\begin{eqnarray}
g^{\rm q}_A=\bigl<2A_3^{({\rm q})3}\bigr>_N&=&
{\rm sgn}(B)N_C\Bigg\{\Bigl<D_{3b}\Bigr>_N
\langle{\rm val}|\sigma_3\tau_b|{\rm val}\rangle
\label{gaq1} \\ &&\hspace{2cm}
+\frac{1-\xi-\eta}{2\alpha^2}\sum_{\mu\ne{\rm val}}
\frac{iT_\mu\epsilon_{3bj}}{\epsilon_{\rm val}-\epsilon_\mu}
\left[\Bigl<D_{3b}J_j\Bigr>_N-
\Bigl<J_jD_{3b}\Bigr>_N\right]\Bigg\},
\nonumber
\end{eqnarray}
where the subscript, $N$, indicates a matrix element between nucleon
states.  Here it should be stressed that the expression in square
brackets vanishes classically; however, in the quantum description
$\mbox{\boldmath $J$}$ does not commute with the rotation matrices
\begin{eqnarray}
[J_i,D_{ab}]=i\epsilon_{ibc}D_{ac}.
\label{comrule}
\end{eqnarray}
Then one finds for (\ref{gaq1})
\begin{eqnarray}
g^{\rm q}_A={\rm sgn}(B)N_C\left\{\Bigl<D_{3b}\Bigr>_N
\langle{\rm val}|\sigma_3\tau_b|{\rm val}\rangle
-\frac{1-\xi-\eta}{\alpha^2}\Bigl<D_{33}\Bigr>_N
\sum_{\mu\ne{\rm val}}
\frac{T_\mu}{\epsilon_{\rm val}-\epsilon_\mu}\right\}.
\label{gaq2}
\end{eqnarray}
The $1/N_C$ correction depends on the parameters $\xi$ and $\eta$
and is hence ambiguous. In various quark soliton models it has been
shown that the prescription $\xi=\eta=0$, which is suggested by the
form of the cranked wave--function (\ref{delval}), results in sizable
corrections to $g_A$ improving the agreement with the experimental
data \cite{wa93}--\cite{ho93}. However, we will now argue that
requiring the proper behavior of the axial current under particle
conjugation symmetry requires us to choose an ordering prescription so
that the matrix elements of the $1/N_C$ terms actually vanish.

The particle conjugation symmetry is most conveniently implemented
as $G$--parity invariance \cite{ni51}. This transformation represents
charge conjugation followed by an isorotation of angle $\pi$ around
the $y$--axis. Under $G$--parity the pion field
changes sign, {\it i.e.}
\begin{eqnarray}
U(\mbox{\boldmath $r$},t)\
{\buildrel{G}\over\longrightarrow}\
U^{\dag}(\mbox{\boldmath $r$},t).
\label{gparu}
\end{eqnarray}
Noting (\ref{rothedge}), we see that $G$--parity reflection for the
rotating soliton corresponds to
\begin{eqnarray}
F(r)\ {\buildrel{G}\over\longrightarrow}\ -F(r)
\label{gparaf}
\end{eqnarray}
while keeping the collective coordinates, $A(t)$ unchanged. At the
soliton level the $G$--parity reflection agrees with the particle
conjugation because the baryon number density reverses its sign under
the transformation (\ref{gparaf}), {\it i.e.} $B\rightarrow-B$. This
property is a consequence of the fact that the isoscalar vector current
has negative $G$--parity. The effect of the transformation (\ref{gparaf})
on the Dirac Hamiltonian may be simplified by introducing the self--adjoint
unitary transformation ${\cal J}=i\beta\gamma_5$ and noting that
\begin{eqnarray}
h(-F)=-{\cal J}^{\dag}h(F){\cal J}.
\label{hrevf}
\end{eqnarray}
{}From this one may read off the particle conjugation properties
of the eigenvalues and eigenstates of the Dirac Hamiltonian
(\ref{hstat}):
\begin{eqnarray}
\epsilon_\mu\ {\buildrel{F\rightarrow -F}
\over\longrightarrow}\ -\epsilon_\mu
\qquad{\rm and}\qquad
|\mu\rangle\
{\buildrel{F\rightarrow -F}\over\longrightarrow}\
{\cal J}|\mu\rangle,
\label{statrev}
\end{eqnarray}
where $\mu$ labels the particular eigenstate. Since ${\cal J}$ does
not affect isopsin it is obvious that the Lagrangian (\ref{alquark})
is invariant under particle conjugation, {\it i.e.} the mass of the
anti--baryon is identical to that of the baryon. Furthermore, as
${\cal J}$ has negative parity, the parity of the anti--baryon is
opposite to that of the baryon, as it should be.
Since the isovector axial vector current $A_\mu^a$, is directly
related to the pion field via $PCAC$ we must have
\begin{eqnarray}
A_\mu^a\ {\buildrel{G}\over\longrightarrow}\ -A_\mu^a ,
\label{gparamua}
\end{eqnarray}
under $G$--parity reflection.

Now we have collected all necessary ingredients and may consider the
effect of particle conjugation on $g_A$. The reduced matrix element
$T_\mu$ in (\ref{sproj}) is invariant under this transformation because
${\cal J}\sigma_i=\sigma_i{\cal J}$. Thus we find\footnote{For taking
matrix elements we may note
$\langle p\uparrow |D_{3b}|p\uparrow\rangle=
\langle p\uparrow |D_{33}|p\uparrow\rangle\delta_{3b}=
\langle \bar{n}\uparrow |D_{33}|
\bar{n}\uparrow\rangle\delta_{3b}=
-1/3\delta_{3b}$.}
\begin{eqnarray}
g^{\rm q}_A{\buildrel{F\rightarrow -F}
\over\longrightarrow}\ \bar{g}^{\rm q}_A=
{\rm sgn}(\bar{B})N_C\left\{\Bigl<D_{3b}\Bigr>_{\bar{N}}
\langle{\rm val}|\sigma_3\tau_b|{\rm val}\rangle
+\frac{1-\xi-\eta}{\alpha^2}\Bigl<D_{33}\Bigr>_{\bar{N}}
\sum_{\mu\ne{\rm val}}
\frac{T_\mu}{\epsilon_{\rm val}-\epsilon_\mu}\right\}
\label{gabar}
\end{eqnarray}
since the energies $\epsilon_\mu$ reverse their signs. The leading
term changes sign under particle conjugation
since ${\rm sgn}(\bar{B})=-{\rm sgn}(B)$. This agrees with
the required transformation property of the axial current under
$G$--parity (\ref{gparamua}).
However, the $1/N_C$ correction does not change sign under
(\ref{statrev}). Stated otherwise, the result for $g_A$ violates the
$G$--parity reflection symmetry unless
\begin{eqnarray}
1-\xi-\eta\equiv0.
\label{equiv0}
\end{eqnarray}
Thus $G$--parity invariance implies an operator ordering which rules
out the $1/N_C$ corrections to $g_A$. We stress that this conclusion
has been obtained without specifying the explicit wave--functions of
the baryon or the anti--baryon. The constraint (\ref{equiv0}) is merely
a consequence of the fact that the leading and next--to--leading
order in $1/N_C$ terms in the approximate expression (\ref{gaq1}) for
$g_A$ transform oppositely under particle conjugation. It is interesting
to note that the hermitean ordering $\xi=\eta=1/2$ satisfies the
constraint (\ref{equiv0}).

Here we have been studying only the contributions of the {\it quarks}
to $g_A$ in the chiral quark model. The leading order term of the
mesonic axial current contains only odd powers of $F(r)$ and
no $1/N_C$ corrections appear. Thus the transformation property
(\ref{gparamua}) is trivially satisfied for the total current
$A_\mu^a$.

Although we have only considered the chiral quark model as the
simplest containing soliton solutions, our studies apply to more
complicated ones as well. In the literature the chiral bag \cite{ho93}
and Nambu--Jona--Lasinio (NJL) \cite{wa93} models have been considered
in connection with the $1/N_C$ corrections to $g_A$. In both models
matrix elements of the structure (\ref{delcur},\ref{sproj}) are involved
in the subleading terms for $g_A$. Since the moment of inertia carries
mass dimension one, the function multiplying the dimensionless reduced
matrix element must be odd in the energy eigenvalues $\epsilon_\mu$.
Thus the corrections do not transform properly under particle conjugation
and an operator ordering which makes these corrections vanish has to
be adopted.

Similar ambiguities appear in the next--to--leading order in the
$1/N_C$ expansion of the isovector part of the magnetic moment
of the nucleon, $\mu_V$. We have also considered the behavior of these
corrections under particle conjugation \cite{sch94}. For this observable
too, the proper behavior requires an ordering so that the $1/N_C$
corrections are zero.

To summarize, we have seen that particle conjugation symmetry
provides useful constraints on operator ordering ambiguities which
may occur when quantizing the chiral soliton. We have seen explicitly
that these constraints require an ordering so that the
next--to--leading $1/N_C$ correction to the axial charge of the
nucleon vanishes in the chiral quark model.

\vskip1.5cm
\leftline{\it Acknowledgements}
We would like to thank R. Alkofer for helpful discussions.
One of us (HW) acknowledges support by a Habilitanden--scholarship
of the Deutsche Forschungsgemeinschaft (DFG).
This work was supported in part by the US DOE contract number
DE-FG-02-85ER 40231.

\vskip1.5cm

\small
\baselineskip=12 true pt

\end{document}